\documentclass[preprint,12pt]{elsarticle}

\usepackage{amssymb}

\begin{document}

\begin{frontmatter}



\title{Refractive index changes of multi-layered spherical nanostructures  with donor
impurity}


\author{Yaghoob Naimi\corref{cor1}}
\ead{y.naimi@iaulamerd.ac.ir}
\author{A. R. Jafari}
\cortext[cor1]{Corresponding author}
\address{Department of Physics, Lamerd Branch, Islamic Azad University, Lamerd, Iran}

\begin{abstract}
In this study, the linear, third-order nonlinear and total refractive index (RI) changes
of multi-layered quantum dot (MLQD) and  multi-layered quantum anti-dot (MLQAD)
with a hydrogenic impurity are calculated. Our numerical results indicate that the RI changes are considerably sensitive to the geometrical parameters of systems and the incident optical intensity. It is observed that by changing the incident photon energy, the RI curves corresponding to MLQAD and MLQD are considerably different in shapes and behaviors. Generally, our results show that MLQAD models have very higher RIs than MLQD models.

\end{abstract}


\end{frontmatter}


\section{Introduction}
\label{}
Spherical-shaped semiconductor nanostructures such as  quantum dots (QDs) and quantum anti-dots (QADs), have gathered a great deal of attention
in the last decades \cite{hei,gru,masu,varshini,sadeghi,vasegh}. These confined structures,
have atomic-like discrete energy levels (subbands) and particular optical properties. Recently, investigations of the physical properties of some new nanostructures such as the multi-layered quantum dots (MLQDs) and anti-dots (MLQADs) have attracted the attention of researchers \cite{cheng,naimi1,karimi1}. These newfound spherical nano-systems can be created by synthesizing different semiconductor materials. The investigation of properties of such systems when they are doped with shallow donor impurities is one of the most studied problems during the last decade \cite{Taş,naimi2,davat,holo}. Impurity existence in such systems plays the essential role in their electrical and optical properties. The nonlinear optical properties (absorbtion coefficient and refractive index) associated with optical absorption of mentioned nanostructures are known to be much stronger than bulk material, due to the quantum confinement effect \cite{atan,tsan,karimi,yang,xie,bas}. These particular optical properties have the highly advantage and potential to construct the optoelectronic devices, such as photo-detector, quantum dot laser and high speed electro-optical modulators \cite{tro,jia,liu,len}. Due to relevance of QDs and QADs to several technological applications, their linear and nonlinear absorption coefficient and refractive index changes have been investigated both theoretically and experimentally by many authors \cite{efo,reza,saf,xi,li,zha,lia,kara,oz1,oz2}.\\

The optical properties of QDs was considered by Efros \cite{efo} for the first time. He studied the light absorption coefficient in a
spherical QD with infinitely height walls. In ref \cite{saf}, the authors have calculated not only the linear and nonlinear absorption
coefficients but also the refractive index changes in a three-dimensional Cartesian coordinate quantum box with finite confining potential
barrier height. The numerical simulations in \cite{Vahdani} show the RI changes in a parabolic cylinder QD and indicate that RI changes are strongly affected by the dot size, the optical intensity and the  polarization of electromagnetic field.
The linear, nonlinear and total refractive index changes and absorption coefficients for transitions
in a spherical QD with parabolic potential have recently been studied in ref \cite{oz1}. Their results, expressed in several allowed
transitions $1s-1p$, $1p-1d$ and $1d-1f$, show that  the transition between orbitals with big $l$ (orbital quantum number) values move to lower incident photon energy region in the presence of parabolic potential term. \\

The outline of this paper is as follows. The theory and
formulation for both MLQAD and MLQD are briefly presented in Section 2. The
linear, nonlinear and total RIs of both models are plotted for various conditions as the function of incident photon energy in Section 3. In this Section the comparison between the behaviour of both models are presented. A brief summary is presented in the last section.\\
\section{Theory and formulation}
\subsection{Hamiltonian and Schr\"{o}dinger equation}
\begin{figure}
\centering
\includegraphics[height=1.89in,width=3.2in]{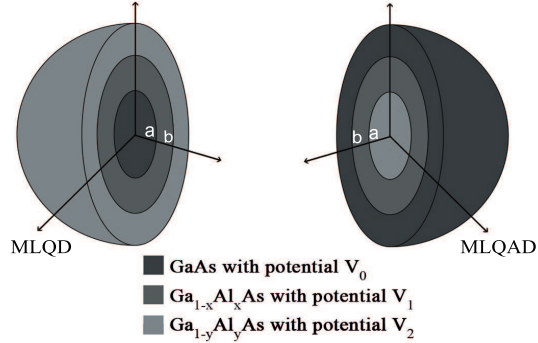}
\caption{A schematic view of the MLQD and MLQAD structures} \label{f1}
\end{figure}

By uniting of GaAs, Ga$_{1-x}$Al$_{x}$As and Ga$_{1-y}$Al$_{y}$As materials, one can make the multi-layered nanostructures. We introduce two types
of these adjacent material connections next to each other as shown in Fig.~\ref{f1}. More precisely, a MLQD (MLQAD) consists of a spherical core made of GaAs (Ga$_{1-y}$Al$_{y}$As) surrounded by a spherical shell
of Ga$_{1-x}$Al$_{x}$As (Ga$_{1-x}$Al$_{x}$As), encompassed in the bulk of Ga$_{1-y}$Al$_{y}$As (GaAs). In other words, the MLQAD is made whenever the core material and the bulk material in MLQD change places. We denote $a$ for the core radius
and $b$ for the total dot (core plus shell) radius, therefore, shell thickness is $b-a$.

Let us consider an electron confined in the MLQD (MLQAD). Within the framework of the effective mass approximation, the Hamiltonian of the system, with hydrogenic impurity in center, is given by
\begin{equation}\label{1} \hat{H} =-\frac{\hbar^{2}}{2\mu(r)}\nabla^{2} -\frac{e^2}{4\pi\varepsilon r}+V(r)\end{equation}
where $ \mu(r)$ and $\varepsilon$ are the electronic effective mass and dielectric constant in the semiconductor medium.  In the different regions, $m(r)$ is represented
as follows

 \begin{equation} \label{2} m(r) =  \left\{
 \begin{array}{ll}
   \mu_{1} & \hspace{1.5cm}r<a \\
   \mu_{2} &  \hspace{1.5cm} a\leq r \leq b\\
  \mu_{3} & \hspace{1.5cm} r>b
 \end{array}\right.
 \end{equation}
$V(r)$ is the confining potential (CP) that corresponding to MLQAD and MLQD has a following form
\begin{equation} \label{2} V^{MLQAD}(r) =  \left\{
\begin{array}{ll}
V_{0} & \hspace{1.5cm}r<a \\
V_{1} & \hspace{1.5cm} a\leq r \leq b\\
V_{2} & \hspace{1.5cm} r>b
\end{array}\right.
\end{equation}
and
\begin{equation} \label{2} V^{MLQD}(r) =  \left\{
\begin{array}{ll}
V_{2} & \hspace{1.5cm}r<a \\
V_{1} & \hspace{1.5cm} a\leq r \leq b\\
V_{0} & \hspace{1.5cm} r>b
\end{array}\right.
\end{equation}
 The barrier height $V_{i}$ arises due to a mismatch between the electronic
affinities of the adjacent media in MLQAD and MLQD, so $V_{1}$ ($V_{2}$) corresponds to binding the Ga$_{1-x}$Al$_{x}$As (Ga$_{1-y}$Al$_{y}$As) to  GaAs (Ga$_{1-x}$Al$_{x}$As). The value of $V_{0}$ is always zero (i.e., the GaAs material has no CP) but writing $V_{0}$
enable us to write next equations in the compact form as you will see.

For many QD heterostructures, such
as GaAs/Ga$_{1-x}$Al$_{x}$As, the image charge effects and polarization
can be notable in the MLQD and MLQAD if there
is a large dielectric discontinuity between the core dot and the
surrounding medium. However, this is not the case for the
GaAs/Ga$_{1-x}$Al$_{x}$As multi-layered systems \cite{Ada}, therefore these effects
can be ignored safely in our calculations. Furthermore, for the sake of generality, the difference between the electronic effective mass
in the dot (antidot) core, the shell and the bulk materials have been ignored (i.e., $\mu_{1}=\mu_{2}=\mu_{3}=\mu$ ). The energies are measured in
$meV$, the effective Rydberg, $Ry=\mu e^4/2\hbar^2(4\pi\varepsilon)^2$, and distances are expressed in $a_{0}=4\pi\varepsilon\hbar^2/(\mu e^2)$. For instance, in the particular case of GaAs-based semiconductors, $\mu = 0.067m_{e}$, and $\varepsilon = 13.18\varepsilon_{0}$.
Thus, for a GaAs host the effective Rydberg and the effective Bohr radius numerically are $Ry = 5.2 meV$ and $a_{0} = 10.4 nm$ respectively. \\

The Schr\"{o}dinger equation is $\hat{H} \psi (r,\theta,\varphi)=E\psi (r,\theta,\varphi)$, by use of separation variable method and this fact that the hydrogenic potential is spherically symmetric, the eigenfunction in the spherical coordinate can be rewritten as $\psi(r,\theta,\varphi)= R(r)Y(\theta,\varphi)$. $Y(\theta,\varphi)$ is given by the spherical harmonics \cite{zet} and it is independent of radial components thus we only focus on the radial part, $R(r)$, when we investigate the optical properties of system. The radial part ($R(r)$) is affected not only by the core and total radius but also by the CPs.

\subsection{The radial Schr\"{o}dinger equation  for MLQAD and MLQD}
The radial part of Schr\"{o}dinger equation, in the spherical coordinate, for both MLQAD and MLQD model takes the form
 \begin{equation}\label{4}
\frac{\hbar^{2}}{2\mu}(\frac{d^{2}}{dr^{2}}+\frac{2}{r}\frac{d}{dr}-\frac{l(l+1)}{r^{2}})R^{(i)}(r)+(E-V_{i}+\frac{e^{2}}{4 \pi \varepsilon r})R^{(i)}(r)=0
 \end{equation}
where $i=2$, $1$ and $0$ ($i=0$, $1$ and $2$) are corresponding to $r<a$, $a\leq r\leq b$ and $r>b$ for MLQAD (MLQD) respectively. Recently in \cite{naimi2}, we have analytically solved this equation for both MLQAD and MLQD and we have shown that the solutions of radial Schr\"{o}dinger equation can be expressed in terms of  Whittaker and hypergeometric functions (One can refer to \cite{naimi2} for reading with full detail) which could be
very complicated, or almost impossible to be applied to the
MLQAD or MLQD to satisfy the boundary conditions. Therefore,
we apply the finite difference method to calculate
the direct numerical solutions for Eq. (5).
\subsection{Linear, nonlinear and total refractive index}
The linear and the third-order nonlinear optical RI changes
for the intersubband transitions can be
calculated by the density matrix approach and the perturbation
expansion method. For
this purpose the system under study can be excited by
an external electromagnetic field of frequency $\omega$, such as the following
\begin{equation}\label{9}
 E(t)=\tilde{E}e^{i\omega t}+\tilde{E}e^{-i\omega t}
  \end{equation}
If $E$ is the perpendicular electromagnetic field along the
$z$ axis, the Hamiltonian of system becomes $H_{0}+ezE(t)$
where where $H_{0}$ is the Hamiltonian of system
without the electromagnetic field $E(t)$. The linear and
the third-order nonlinear optical RI of
a spherical nano-system, within a two-level system, in a
special case, from ground ($\psi_{100}$) to first allowed excited state ($\psi_{210}$) can be expressed as
\cite{saf,zha,kho}.
\begin{equation}
\hspace{0cm} \frac{\Delta n^{(1)}(\omega)}{n_{r}}={\frac{\sigma_{v}|M_{21}|^2}{2n^{2}_{r}\varepsilon_{0}}} \frac{(E_{21}-\hbar\omega)}{(E_{21}-\hbar\omega)^2+(\hbar\Gamma_{21})^2} \end{equation}
\begin{equation}\label{3}
\frac{\Delta n^{(3)}(\omega)}{n_{r}}=
-{\frac{\mu Ic\sigma_{v}|M_{21}|^4}{n^{3}_{r}\varepsilon_{0}}} \frac{(E_{21}-\hbar\omega)}{[(E_{21}-\hbar\omega)^2+(\hbar\Gamma_{21})^2]^2} \end{equation}
Then the total RI is
\begin{equation}
\frac{\Delta n_{r}(\omega)}{n_{r}}=\frac{\Delta n^{(1)}_{r}(\omega)}{n_{r}}+\frac{\Delta n^{(3)}_{r}(\omega)}{n_{r}}
\end{equation}
In above equation we have used
\begin{eqnarray}
&&E_{21}=E_{2}-E_{1},\nonumber \\
&&\varepsilon_{R}=n^{2}_{r}\varepsilon_{0}, \nonumber \\
&&M_{21}=|<\psi_{210}|ez|\psi_{100}>|,
\end{eqnarray}
where $E_{21}$ denotes the difference
of the energy between two lowest electronic
states, $M_{21}$ is an element of electric dipole moment matrix,  $\mu$, $\sigma_{v}$ and $n_{r}$ are the permeability,  carrier density and refractive index of the system respectively. $\hbar\omega$ is the incident photon energy, $\Gamma_{21}=1/T_{21}$ where $T_{21}$ is the phenomenological relaxation rate, caused by the electron-phonon, electron–electron and other collision processes. $I$ is the optical intensity of incident wave, $c$ and $\varepsilon_{0}$ is the speed of light in the free space and permeability of vacuum.
\section{Results and discussion}

In the following, we calculate  the linear, third-order nonlinear and total refractive index changes in both nano-systems with different shell thicknesses and CPs. The unchanged parameters used in our calculations are: $\Gamma_{21}=0.2ps$, $\sigma_{v}=3.0\times10^{22}m^{-3}$, $n_{r}=3.2$, $\varepsilon_{R}=13.18$. \\

According to Eq. (7) and (8), the linear and the nonlinear RI always have an opposite sign and their values depend on the value of $E_{21}$. More exactly, in $\hbar\omega=E_{21}$ both terms are zero and change their sign as $\hbar\omega<E_{21}$ ($\hbar\omega>E_{21}$) the linear term is positive (negative) and the nonlinear term is negative (positive). Furthermore, the smaller $E_{21}$ leads to the larger linear and nonlinear values. In other words, the smaller $E_{21}$ causes the larger value of RI terms and also for smaller $E_{21}$, the RI terms reach zero value in smaller incident photon energy.

\begin{figure}
\mbox{\includegraphics[ height=2.1in,width=2.72in]{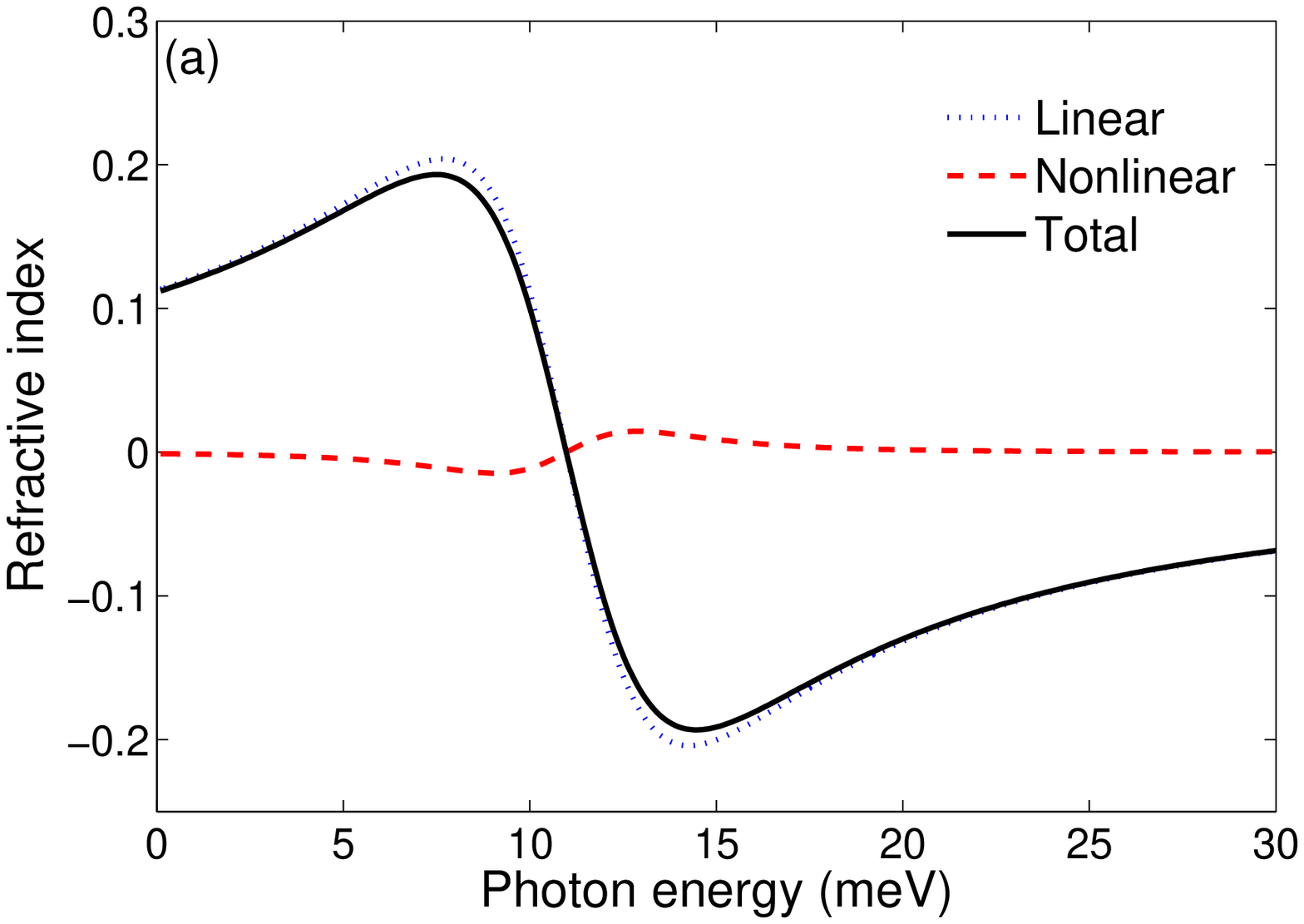}\quad
\includegraphics[ height=2.1in,width=2.72in]{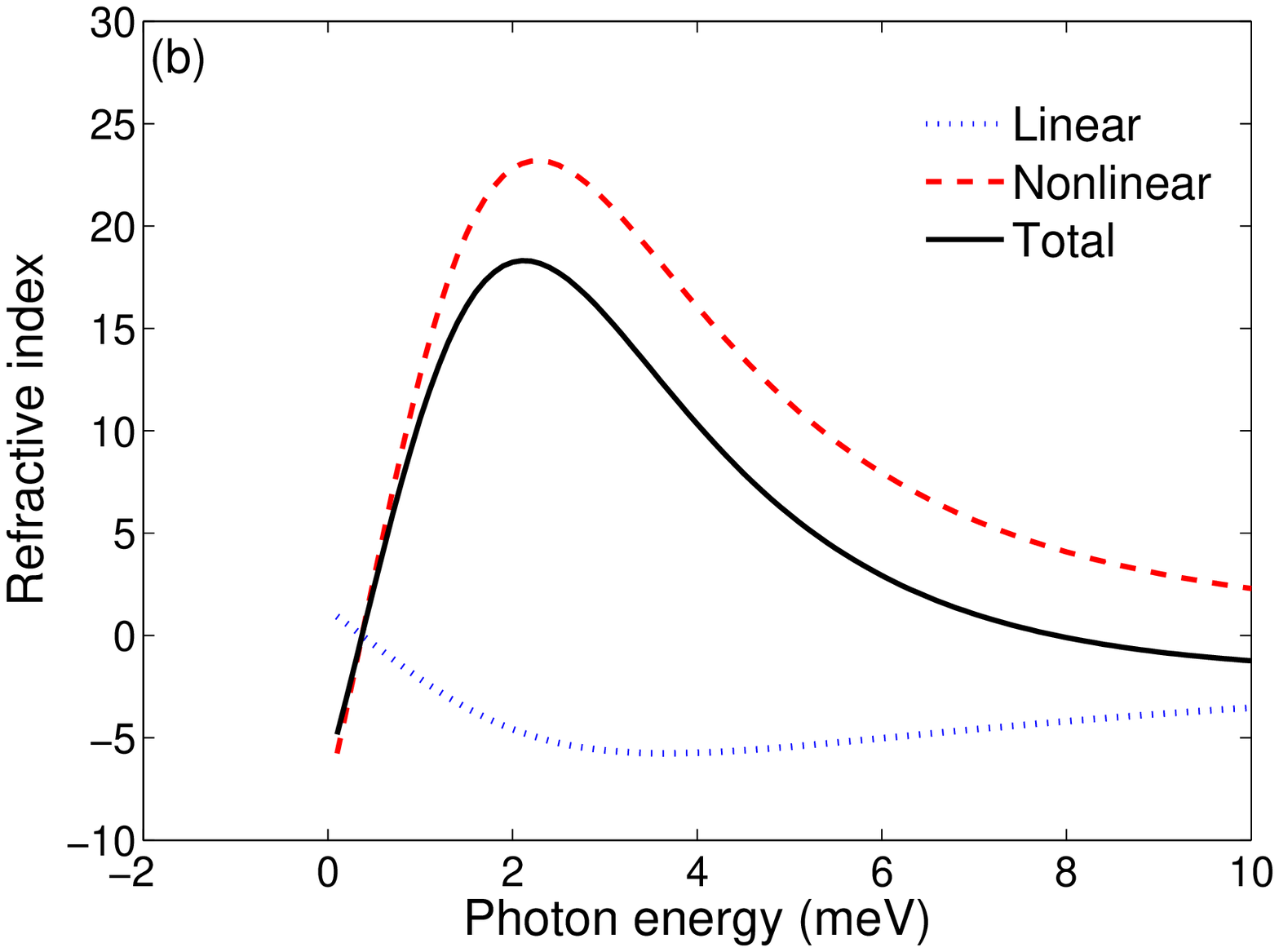} }
 \caption{The linear, nonlinear and total RIs as a function of incident photon energy in the same condition $\{I=200MW/m^{2}$, $b-a=0.5a_{0}$, $V_{1}=2Ry$ and $V_{2}=5Ry\}$ for (a) MLQD (b) MLQAD}
 \label{f2}\end{figure}

Under condition \{$I=200MW/m^{2}$, $a=2a_{0}$, $b-a=0.5a_{0}$, $V_{1}=2Ry$ and $V_{2}=5Ry$\}, our numerical calculations result that the values of $E_{21}$ for nano-systems are $E_{21}^{MLQAD}=0.37  meV$ and $E_{21}^{MLQD}=10.98  meV$, thus we have
\vspace{0.5cm}
\begin{equation}\label{4}
  E_{21}^{MLQAD}\ll E_{21}^{MLQD}
\end{equation}
According to discussion in above discussions, one can conclude that the value of linear and nonlinear RI for MLQAD should be very larger than corresponding value in MLQD and also linear and nonlinear RI for MLQAD should reach zero value in very smaller incident photon energy than MLQD as Fig.~\ref{f2} shows well. These figures indicate that in MLQAD (MLQD) the major and minor terms of total RI are nonlinear (linear) and linear (nonlinear) terms respectively.

\begin{figure}
\mbox{\includegraphics[ height=2.1in,width=2.7in]{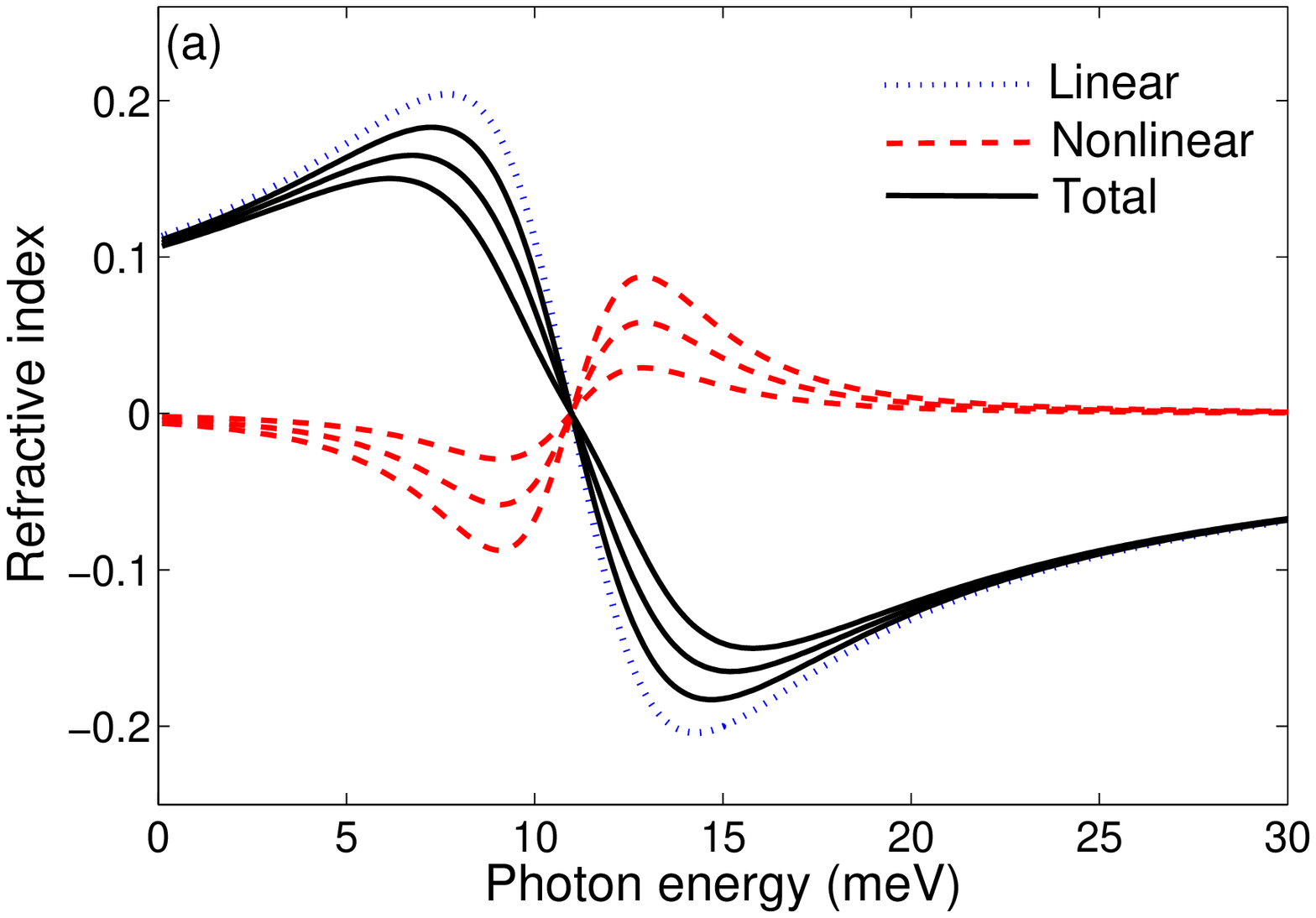}\quad
\includegraphics[ height=2.1in,width=2.7in]{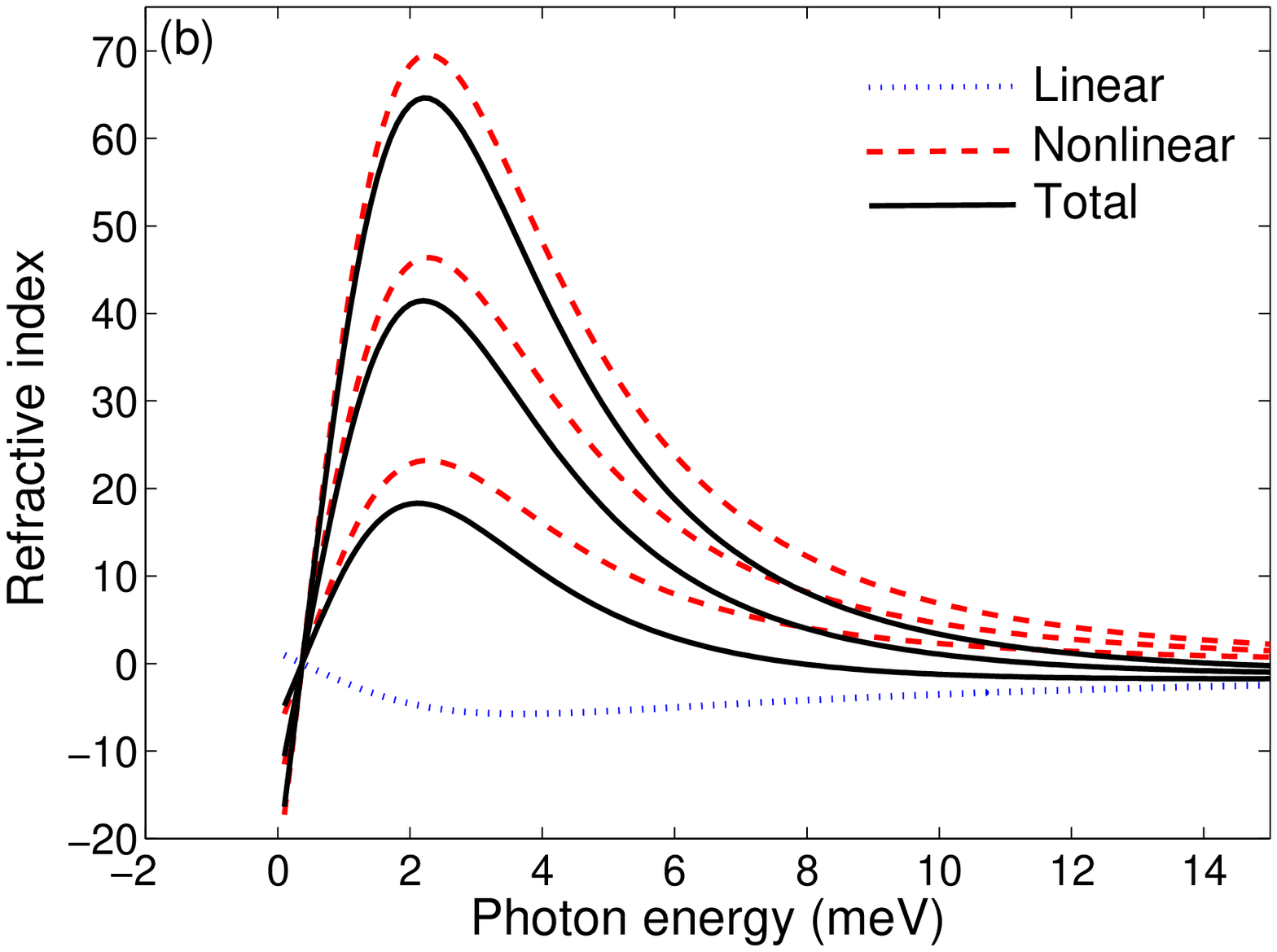} }
 \caption{The linear, nonlinear and total RIs as a function of incident photon energy with the fixed parameters $\{a=2a_{0}, b-a=0.5a_{0},V_{1}=2Ry$ and $V_{2}=5Ry\}$ and three different intensities for (a) MLQD that from up (b) MLQAD that from bottom the intensity value is $I=200MW/m^{2}$, $400MW/m^{2}$ and $600MW/m^{2}$ respectively. }
 \label{f3}\end{figure}

From relation (7)-(9), by increasing the optical intensity the major term (nonlinear term) of total RI of MLQAD increases although the minor term remains unchange,  so we predict that the total RI for MLQAD should be enhanced. But in the case of MLQD, by increasing the optical intensity the major term (linear term) of total RI do not change although the minor term increases. Since the opposite sign of major and minor terms, we predict that by increasing the optical intensity the total RI should be decreased. Fig.~\ref{f3} verifies our predictions.

\begin{figure*}
\mbox{\includegraphics[ height=2.1in,width=2.7in]{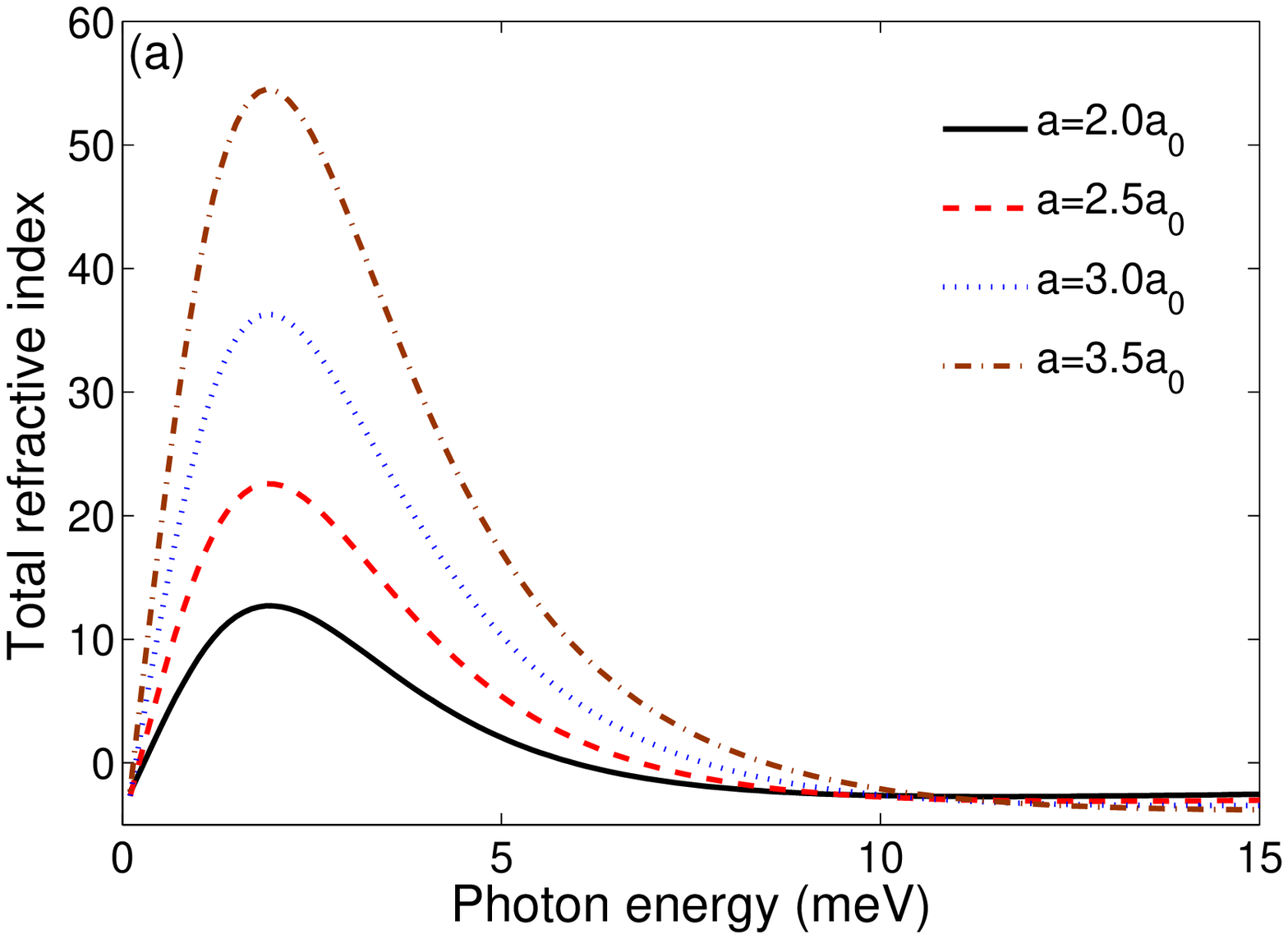}\quad
\includegraphics[ height=2.1in,width=2.7in]{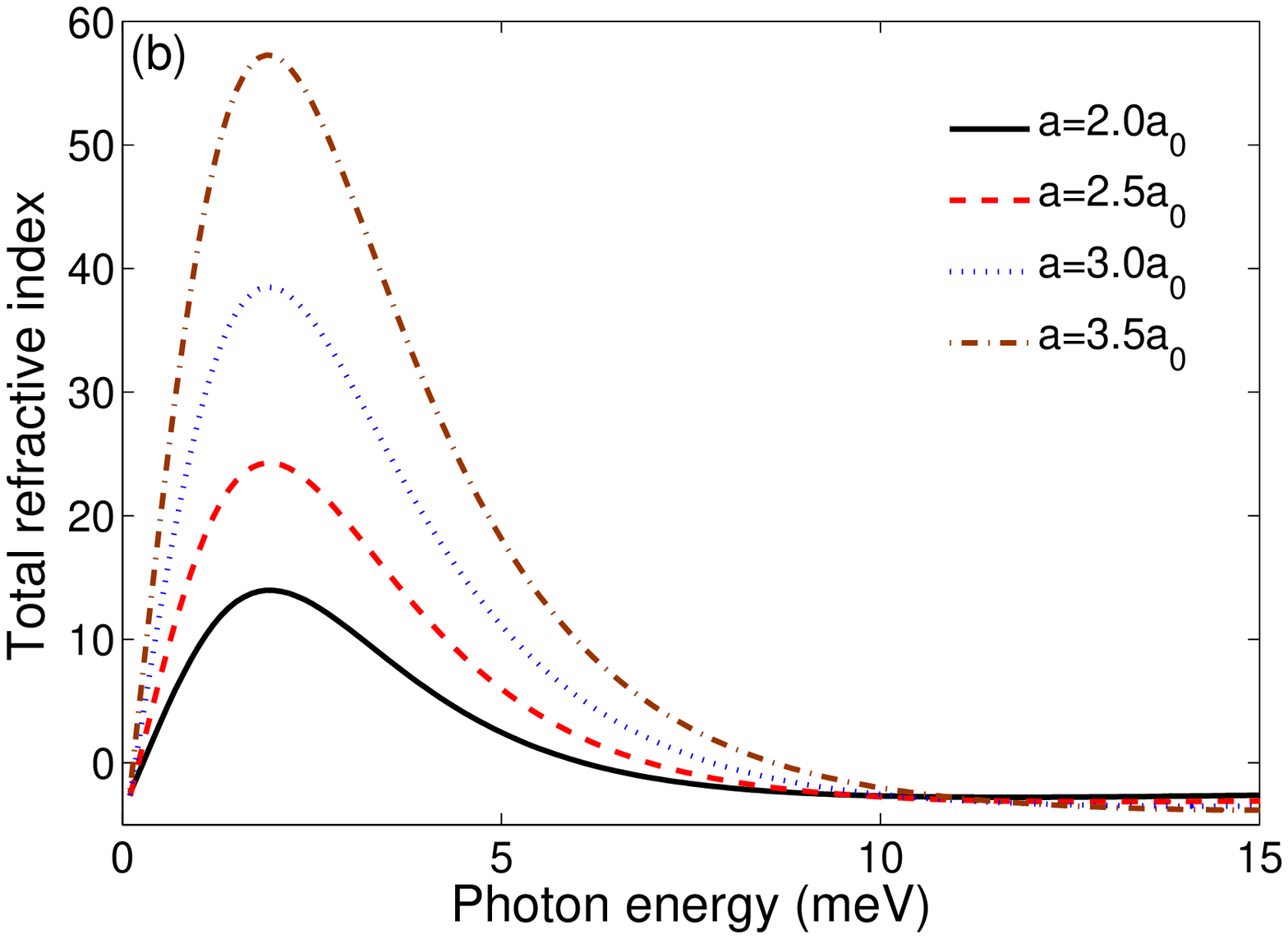} }
 \caption{The total RI of  MLQAD as a function of incident photon energy with the fixed shell thickness $b-a=1a_{0}$ and four different core radii for (a) $V_{1}=2Ry$ and $V_{2}=5Ry$ (b) $V_{1}=2Ry $ and $V_{2}=\infty Ry$}
 \label{f4}\end{figure*}

In Figs.~\ref{f4}, the total RIs of MLQAD are plotted when the shell thickness is fixed as $b-a=1a_{0}$ whiles a core radii have a four different values as $a=2, 2.5, 3, 3.5 a_{0}$. In Figs.~\ref{f4} (a) and (b) the values of CPs are \{$V_{1}=2Ry, V_{2}=5Ry$\} and \{$V_{1}=2Ry, V_{2}=\infty Ry$\} respectively. It is observed that by increasing the core dot radius, the position of  total RI peak has no significant shift but the height of total RI peak enhanced. Furthermore, by changing the value of $V_{2}$, no considerable differences are observed.
\begin{figure*}
\mbox{\includegraphics[ height=2.1in,width=2.72in]{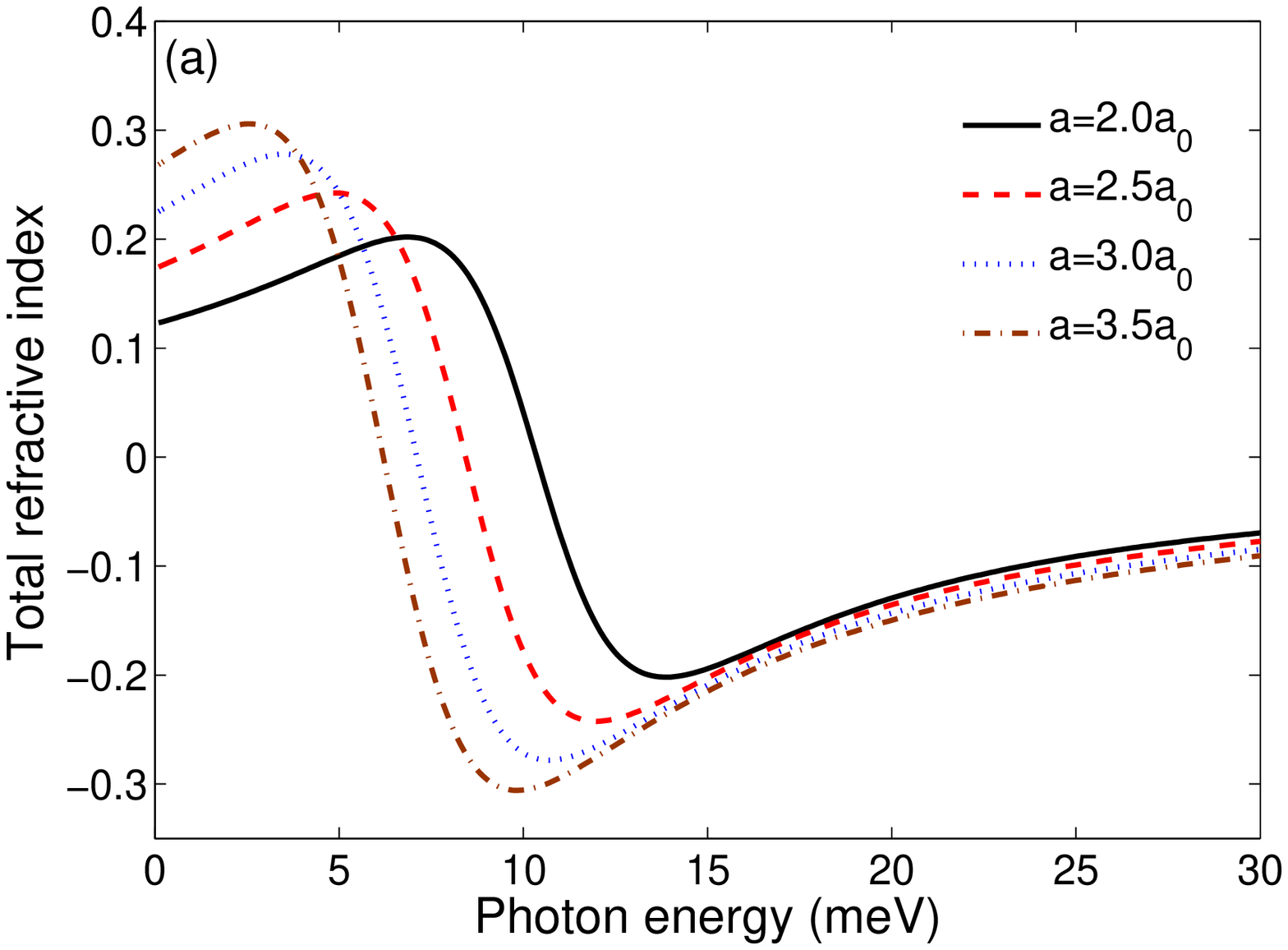}\quad
\includegraphics[ height=2.1in,width=2.72in]{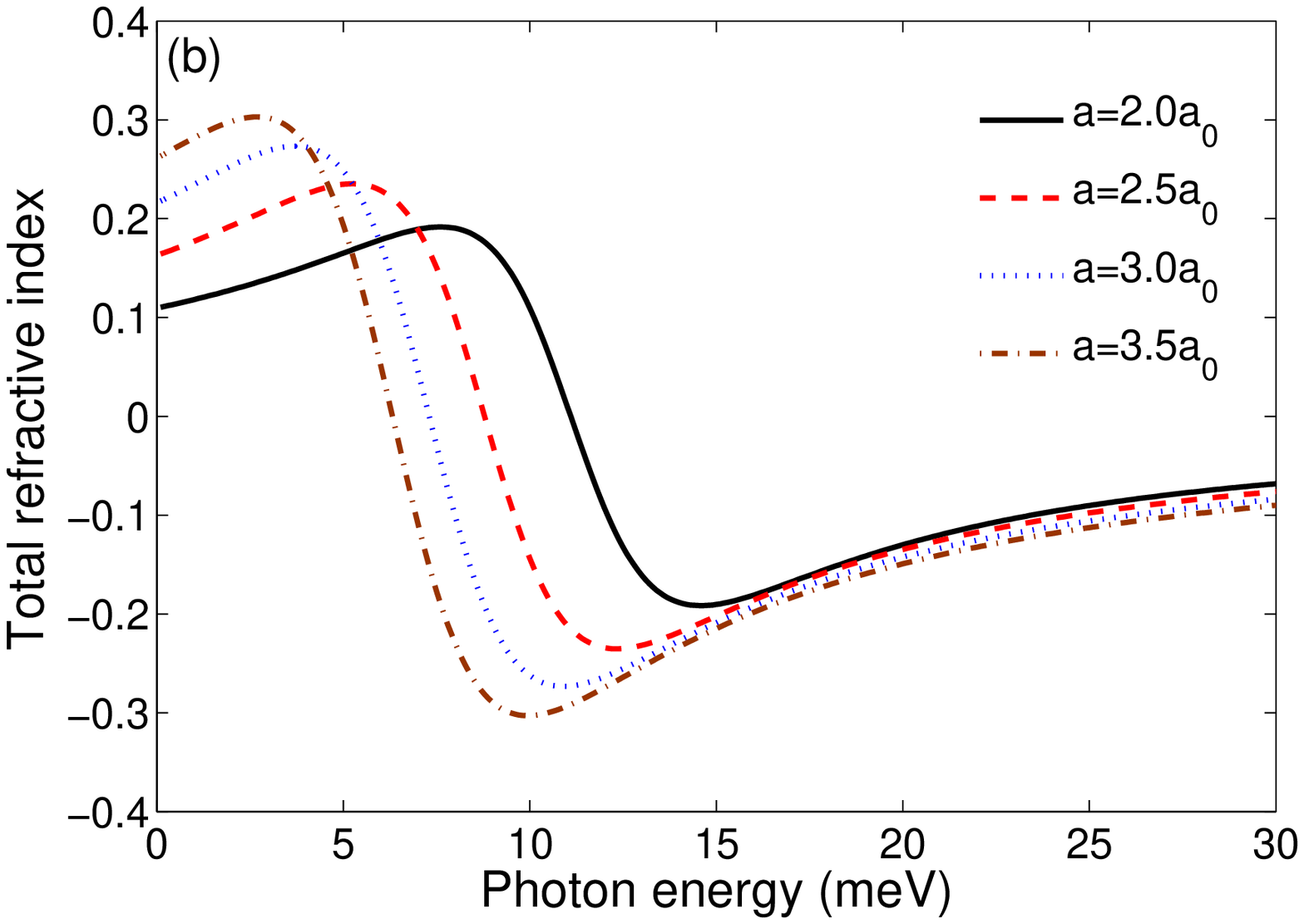} }
 \caption{The total RI of  MLQD as a function of incident photon energy with the fixed shell thickness $b-a=1a_{0}$ and four different core radii for (a) $V_{1}=2Ry$ and $V_{2}=5Ry$ (b) $V_{1}=2Ry $ and $V_{2}=\infty Ry$}
 \label{f5}\end{figure*}

The same conditions are employed for MLQD in Figs.~\ref{f5}. In this case, by increasing the core dot radius, the maximum value of RI becomes bigger and also the peak position shifts toward the lower photon energies. This behaviour arises from this fact that by increasing the core dot radius, the energy difference between ground and first excited states becomes smaller. But in the case of MLQAD (Fig.~\ref{f4}) the energy difference is approximately negligible, thus the peak position remained unchanged. Simultaneously comparison of Fig.~\ref{f5} (a) and (b) shows that RI value not affected by changing in the second potential barrier ($V_{2}$).
\begin{figure*}
\mbox{\includegraphics[ height=2.1in,width=2.72in]{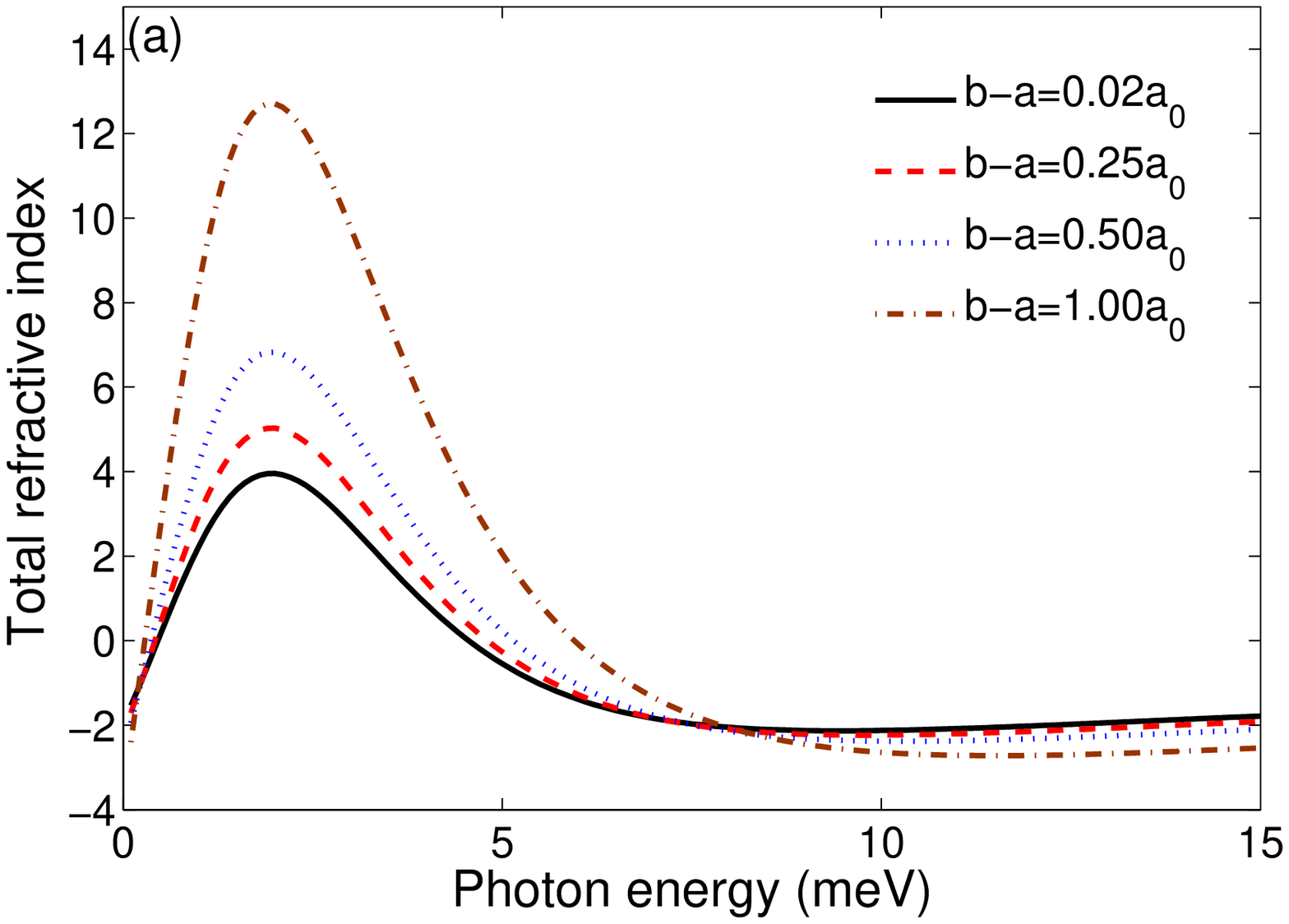}\quad
\includegraphics[ height=2.1in,width=2.72in]{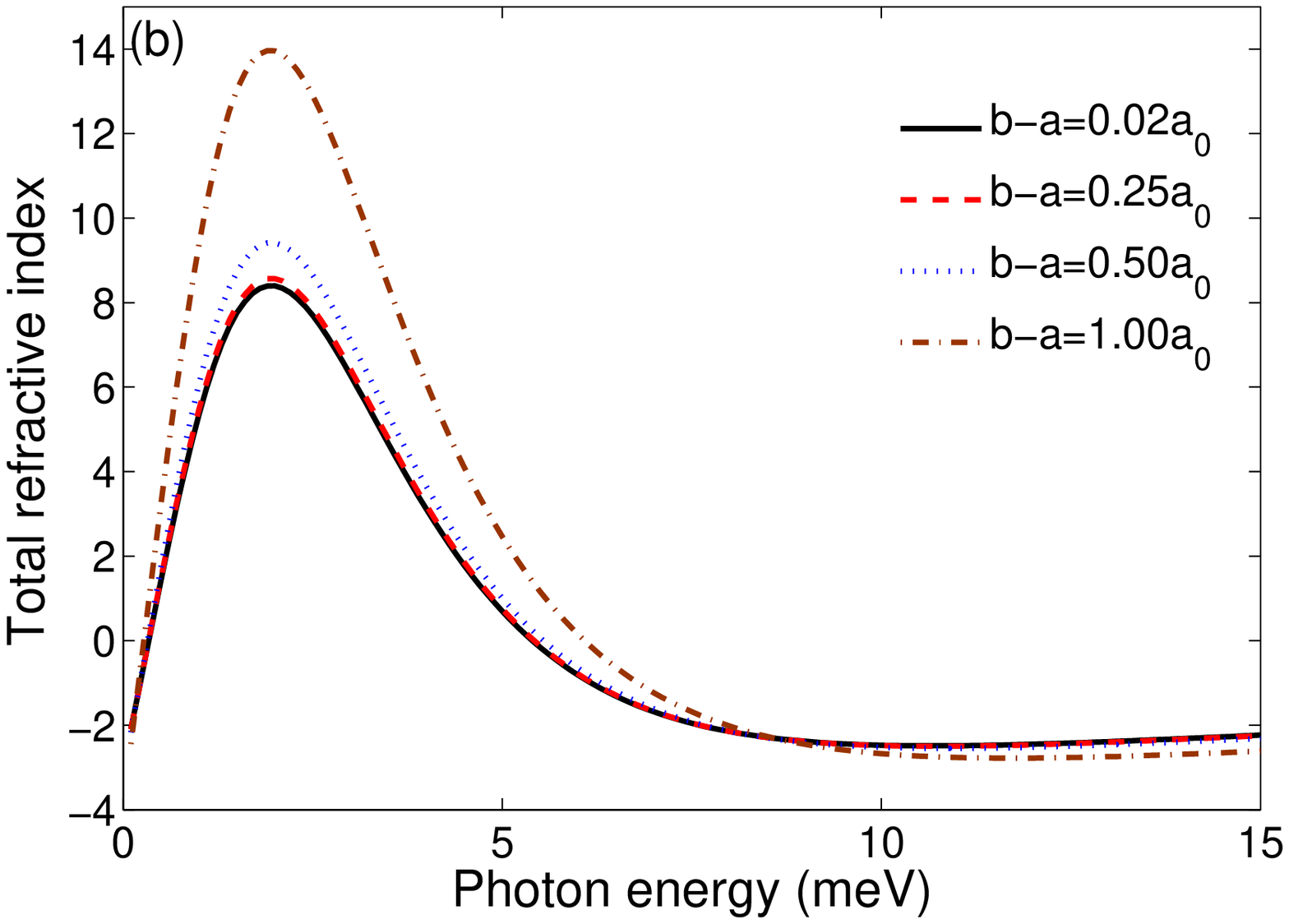} }
 \caption{The total RI of MLQAD as a function of incident photon energy with $a=2a_{0}$ and four different shell thicknesses for (a) $V_{1}=2Ry$ and $V_{2}=5Ry$ (b) $V_{1}=2Ry $ and $V_{2}=\infty Ry$}
 \label{f6}\end{figure*}
\begin{figure*}
\mbox{\includegraphics[ height=2.1in,width=2.72in]{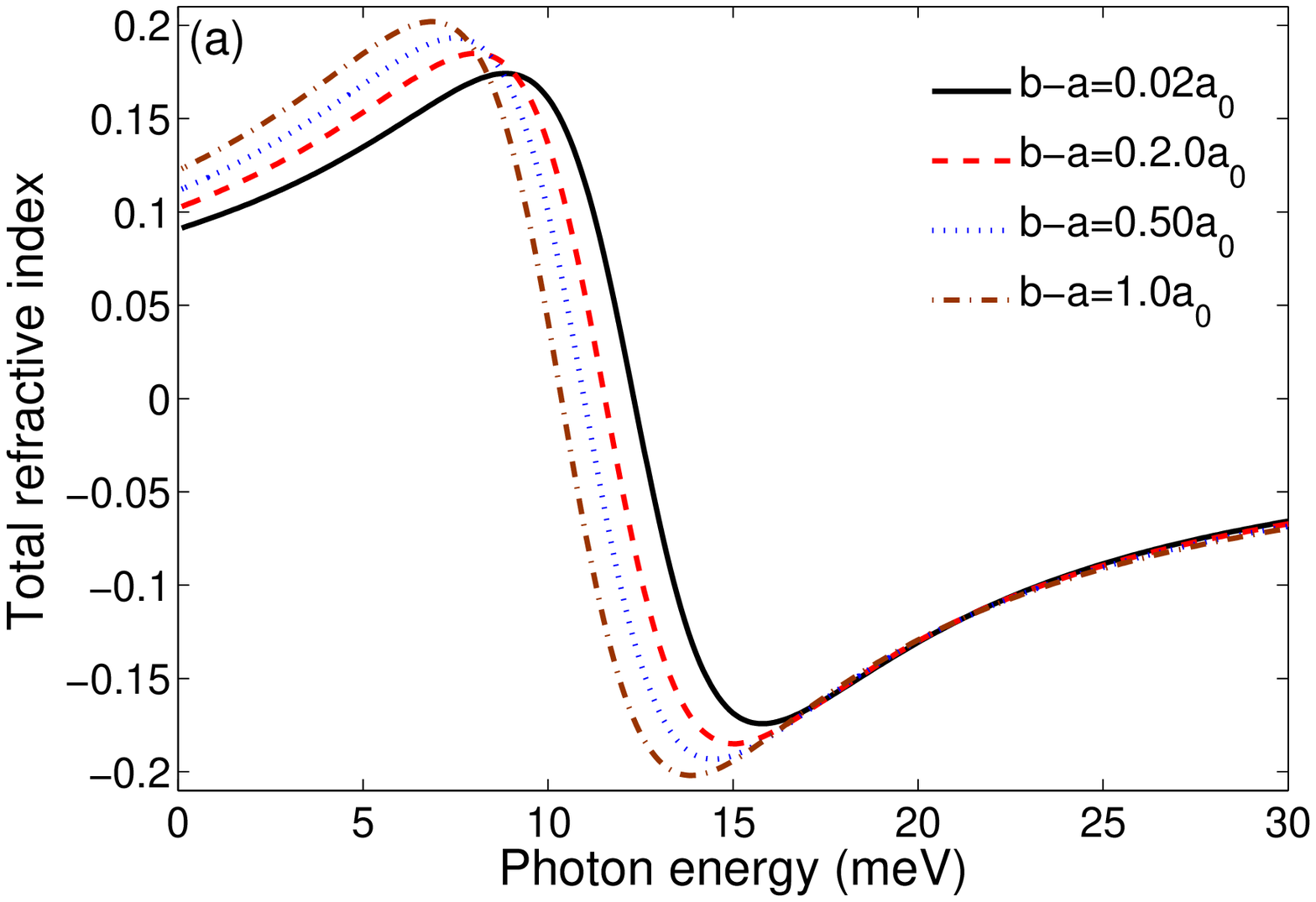}\quad
\includegraphics[ height=2.1in,width=2.72in]{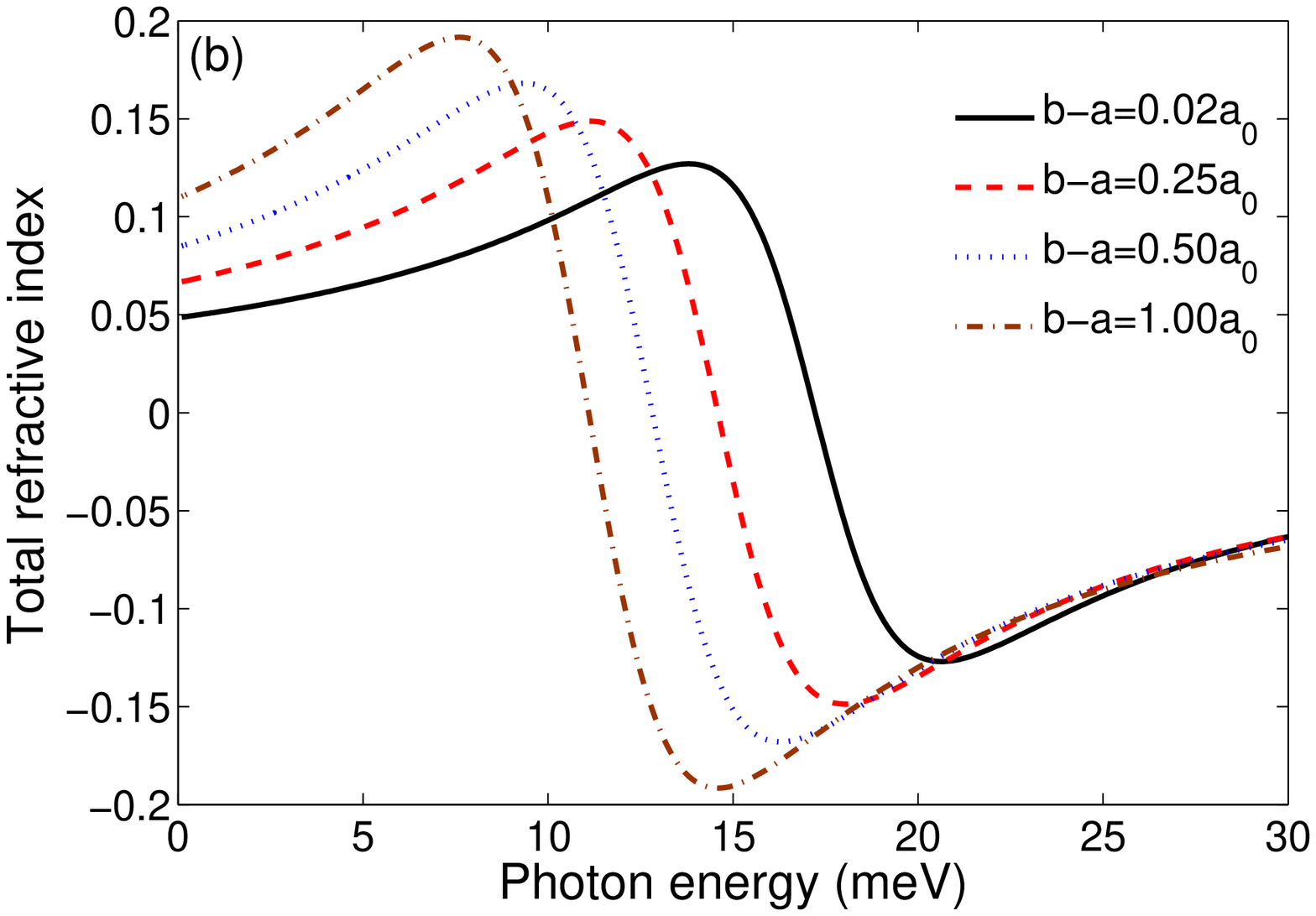} }
 \caption{The total RI of MLQD as a function of incident photon energy with $a=2a_{0}$ and four different shell thicknesses for (a) $V_{1}=2Ry$ and $V_{2}=5Ry$ (b) $V_{1}=2Ry $ and $V_{2}=\infty Ry$}
 \label{f7}\end{figure*}

The total RI curves of both nano-systems for the fixed core radius $a=2.0a_{0}$ and four different shell thicknesses $b-a=1, 0.5, 0.25, 0.02 a_{0}$, as the function of photon energy are presented in Fig.~\ref{f6} and Fig.~\ref{f7}, respectively. The difference between (a) and (b) parts in these figures is the value of $V_{2}$ that is $5Ry$ and infinity respectively. In Fig.~\ref{f6} and Fig.~\ref{f7}, one can see that by increasing the shell thickness, the RI value corresponding to both nano-systems becomes greater. It is observed that by increasing the shell thickness, the peak position of total RI curves for MLQAD approximately remains unchange, however in MLQD the peak and valley of total RI shift to smaller photon energies. The influence of second CP ($V_{2}$) on total RI values is more prominent in the case of different shell thickness rather than constant shell thickness. In Fig.~\ref{f6} and Fig.~\ref{f7}, comparing the curves for different $V_{2}$ within certain shell thickness for both nano-system show that the changes corresponding to smaller shell thickness is more clear rather than others.
\section{Conclusion}
We have investigated the linear, the third-order nonlinear and the total optical RIs of multi-layered spherical nano-systems with a hydrogenic impurity in the center. As our results indicate, the optical RIs changes of MLQAD and MLQD are very different. It is found that in the same conditions MLQAD has very higher RI changes than MLQD. It is observed that for the fixed
shell thickness, by increasing the amount of core radius, the
total RI peak heights become larger and shift toward
lower incident photon energies for MLQD but the
peak heights corresponding to MLQAD become larger with no significant
changes in peaks positions. Also, in this case the, changes in $V_{2}$ value lead to no considerable
changes in total RI curves. Furthermore, for a fixed
core radius, by increasing the shell thickness, the peak of
total RI curves corresponding to MLQAD, moves to larger
values without considerable changes in the incident photon
energy but the peak heights related to MLQD become larger and shift to smaller energy
regions. Finally, it is found that in the latter case, by decreasing the
shell thicknesses, the total RI curves are considerably affected
by $V_{2}$ value.

\end{document}